# Optical Frequency Comb Fourier Transform Spectroscopy of $^{14}N_2^{16}O$ at 7.8 μm

Adrian Hjältén[1], Matthias Germann[1], Karol Krzempek[2], Arkadiusz Hudzikowski[2], Aleksander Głuszek[2], Dorota Tomaszewska[2], Grzegorz Soboń[2], and Aleksandra Foltynowicz[1,*]

[1]*Department of Physics, Umeå University, 901 87 Umeå, Sweden*
[2]*Laser & Fiber Electronics Group, Faculty of Electronics, Wrocław University of Science and Technology, 50-370 Wrocław, Poland*
[*]*aleksandra.foltynowicz@umu.se*

**Abstract:**

We use a Fourier transform spectrometer based on a compact mid-infrared difference frequency generation comb source to perform broadband high-resolution measurements of nitrous oxide, $^{14}N_2^{16}O$, and retrieve line center frequencies of the $\nu_1$ fundamental band and the $\nu_1 + \nu_2 - \nu_2$ hot band. The spectrum spans 90 cm$^{-1}$ around 1285 cm$^{-1}$ with a sample point spacing of $3 \times 10^{-4}$ cm$^{-1}$ (9 MHz). We report line positions of 72 lines in the $\nu_1$ fundamental band between P(37) and R(38), and 112 lines in the $\nu_1 + \nu_2 - \nu_2$ hot band (split into two components with e/f rotationless parity) between P(34) and R(33), with uncertainties in the range of 90-600 kHz. We derive upper state constants of both bands from a fit of the effective ro-vibrational Hamiltonian to the line center positions. For the fundamental band, we observe excellent agreement in the retrieved line positions and upper state constants with those reported in a recent study by AlSaif *et al.* using a comb-referenced quantum cascade laser [J Quant Spectrosc Radiat Transf, 2018;211:172-178]. We determine the origin of the hot band with precision one order of magnitude better than previous work based on FTIR measurements by Toth [http://mark4sun.jpl.nasa.gov/n2o.html], which is the source of the HITRAN2016 data for these bands.

# 1. Introduction

Nitrous oxide ($N_2O$) is of significant importance for atmospheric physics and chemistry. Despite being only a minor constituent of the Earth's atmosphere, it is a potent greenhouse gas [1], and it contributes strongly to stratospheric ozone depletion [2]. Moreover, global $N_2O$ emissions are increasing due to human activity [3] and will likely be enhanced by global warming [4]. This calls for the ability to monitor atmospheric $N_2O$ by spectroscopic means, which requires laboratory studies to provide precise and accurate line parameters. One wavelength range of interest for environmental monitoring is the atmospheric window around 8 μm, where $N_2O$ absorbs due to its strong $v_1$ fundamental band as well as several weaker hot and overtone bands. The 8 μm spectral range has been extensively studied using conventional Fourier transform infrared spectroscopy (FTIR). Guelachvili [5] determined line positions of the $v_1$ band, the $2v_2$ band, and several hot bands with uncertainties of one to a few MHz, as well as relative line intensities and linewidths. Levy *et al.* [6] reported absolute line intensities of the $v_1$ and the $2v_2$ bands, while Lacome *et al.* [7] studied self- and $N_2$-broadening of lines belonging to these two bands. Toth [8-11] measured various $N_2O$ absorption bands over a wide range of the infrared spectrum and compiled an extensive line list [12], including line positions, intensities, and broadening parameters, which to this day forms the basis for a large part of the entries on $N_2O$ in the HITRAN database [13]. More recently, FTIR was employed by Wang *et al.* [14] to measure and assign the spectra of the less abundant $N_2O$ isotopologues around 8 μm. In an early laser-based spectroscopic study of $N_2O$, Varanasi and Chudamani [15] used a tunable diode laser to measure the line intensities of the $v_1$ band. A few years later, Maki and Wells [16] determined frequencies of the $N_2O$ absorption lines in several bands spread over almost the entire infrared spectrum (among them the $v_1$ band, $2v_2$ band and hot bands at ~8 μm) with uncertainties in the single MHz range using heterodyne frequency measurements, for use in infrared frequency calibration tables. As quantum cascade lasers (QCLs) became available, Grossel *et al.* [17] used a continuous-wave (CW) QCL to measure line intensities and air broadening parameters of the $v_1$ band, while Tonokura *et al.* [18] measured $CO_2$ broadening parameters of the $v_1$ band. Moreover, Tasinato *et al.* [19] employed a pulsed QCL to study collisional processes, and hence pressure broadening, of the $v_1$ band in He, Ar, $N_2$, and $CO_2$.

The precision and accuracy of line position determination can be significantly improved by using optical frequency combs that directly link the optical frequencies to radio frequency standards. To this date, there exists only one set of studies of line positions of $N_2O$ in the 8 μm range involving a frequency comb. Lamperti *et al.* [20] referenced a CW QCL tunable around ~8 μm to a Tm:fiber comb at 1.9 μm through a sum frequency generation scheme, and reported the positions of three lines in the $v_1$ band with ~63 kHz uncertainty. AlSaif *et al.* [21] used this comb-referenced QCL to retrieve positions of the P(40) to R(31) lines of the $v_1$ band in the Doppler limit with uncertainties below 200 kHz. More recently, they measured the center frequency of the R(16) line with precision below 50 kHz using sub-Doppler spectroscopy [22]. We show that similar performance (in the Doppler limit) can be obtained by employing an 8 μm comb to directly measure the spectrum, removing the need for complex frequency referencing. Moreover, direct frequency comb spectroscopy allows measuring the entire vibrational bands with tens of thousands of comb lines simultaneously (rather than sequentially line by line, as is done with CW lasers), which reduces the influence of drifts. The most promising comb sources for precision spectroscopy in the 8 μm range are based on difference frequency generation (DFG) from a single femtosecond pump source, deriving the signal comb from the pump laser using a nonlinear fiber [23-25], or through intrapulse DFG (IDFG) [26, 27]. These DFG combs cover wide bandwidths (up to superoctaves for IDFG sources), and are inherently free from carrier-envelope-offset frequency, $f_{ceo}$, since $f_{ceo}$ is identical for the pump and signal combs and cancels in the DFG process. The lack of $f_{ceo}$ significantly simplifies the absolute stabilization of the DFG-based comb sources. Timmers *et al.* [26] employed IDFG sources for dual-comb spectroscopy (DCS) with sample point spacing equal to the comb repetition rate, $f_{rep}$, of 100 MHz. Much denser sampling point spacing was demonstrated by Changala *et al.* [28], who used a DFG source [24] and a Fourier transform spectrometer (FTS) with comb-mode-width limited resolution to measure the ro-vibrational spectrum of $C_{60}$. The 8 μm range can also be reached through nonlinear conversion in optical parametric oscillators (OPOs), which, however, are not $f_{ceo}$-free. 8 μm OPOs have been combined with an FTS [29], the DCS approach [30], and an immersion grating spectrometer [31], all with resolution limited by the instrumental broadening rather than the comb mode linewidth. The latter work investigated the pressure broadening of one hot band $N_2O$ line, but does not report the line positions. It should be noted that frequency combs based on QCLs [32] operate in the 8 μm range. However, their inherently large comb mode spacing is not ideal for precision spectroscopy of molecules in gas phase at low pressures. The recently demonstrated interleaving method [33] overcomes this problem; however, for absolute frequency calibration, QCL combs still require an external optical reference in the form of a mode-locked comb [34].

We use a Fourier transform spectrometer and a recently developed compact all-fiber-based $f_{ceo}$-free DFG comb emitting at 8 μm [25] to perform high-resolution absorption measurements of the $\nu_1$ fundamental band and the $\nu_1 + \nu_2 - \nu_2$ hot band of $^{14}N_2^{16}O$. We retrieve line positions of 72 absorption lines of the $\nu_1$ band between 1250 and 1315 cm$^{-1}$. By means of a least-square fit of an effective ro-vibrational Hamiltonian to these line positions, we obtain values for the vibrational term value $G_v$, the rotational constant $B_v$, and the quartic and sextic centrifugal distortion constants $D_v$, $H_v$ of the $\nu_1$ vibrational excited state. Moreover, we retrieve line positions of 112 absorption lines belonging to the $\nu_1 + \nu_2 - \nu_2$ hot band. From a fit of the effective Hamiltonian, we obtain the band origin and the $B_v$ and $D_v$ constants of the e- and f-components of the $\nu_1 + \nu_2$ vibrational excited state of $N_2O$.

## 2. Experimental setup and procedures

The main building blocks of the experimental setup (shown in Fig. 1) are a DFG-based optical frequency comb, a multi-pass absorption cell, and a Fourier transform spectrometer.

The mid-infrared comb source has been described in detail previously in Ref. [25]. In brief, it consists of two parts, an all-fiber femtosecond dual-wavelength source and a free-space module with a nonlinear crystal for difference frequency generation. The dual-wavelength source is based on a thermally stabilized Er:fiber oscillator and emits two femtosecond pulse trains: one around 1.56 μm derived directly from the oscillator, and the other around 1.95 μm generated through the Raman soliton self-frequency shift in a highly nonlinear fiber. The nominal pulse repetition rate, $f_{rep}$, of 125 MHz, can be tuned over a range of ~3 kHz using a piezo (PZT) fiber stretcher. Both pulse trains are guided in a single optical fiber to a fiber-coupled achromatic collimator and focused in free space on an orientation-patterned gallium phosphide (OP-GaP) crystal (BAE Systems) with a poling period of 60 μm. The DFG process taking place in the crystal produces an $f_{ceo}$-free mid-infrared (MIR) optical frequency comb with a spectrum spanning from 1240 to 1330 cm$^{-1}$ and an output power of 1.5 mW. To avoid beam shape fluctuations due to thermal lensing, the crystal is kept at a constant temperature of 40 °C using a temperature-controlled oven (Covesion, PV10).

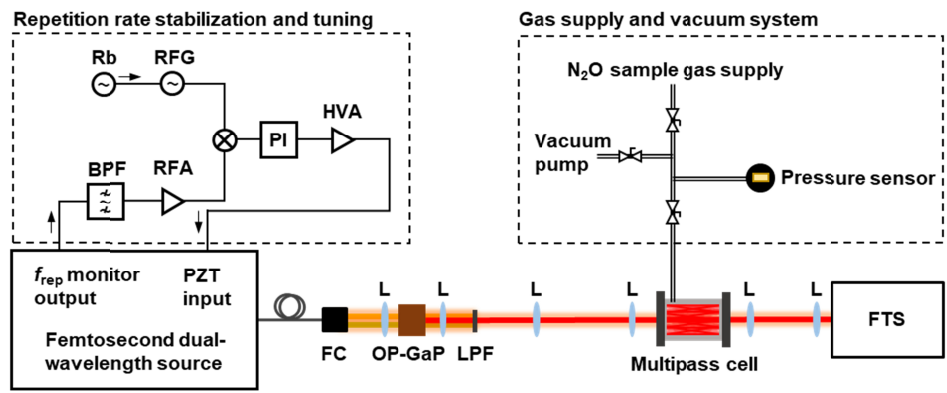

Fig. 1 Schematic of the experimental setup. Rb: rubidium frequency standard, RFG: radiofrequency signal generator, PI: proportional-integral controller, HVA: high-voltage amplifier, BPF: electric bandpass filter, RFA: radiofrequency amplifier, FC: fiber collimator, L: spherical lens, OP-GaP: orientation-patterned gallium phosphide crystal, LPF: optical long-pass filter, FTS: Fourier transform spectrometer.

The repetition rate is stabilized through a phase lock to a tunable RF source (shown in the upper left part of Fig. 1). To this end, a photodiode detects a part of the Er:fiber oscillator output. The 4$^{th}$ harmonic of the photodiode signal is isolated using an RF bandpass filter, amplified, and fed to a double-balanced mixer (all from Minicircuits) for phase comparison with a reference signal synthesized using a low-noise RF generator (Anritsu, MG3692A), which in turn is referenced to a GPS-disciplined rubidium frequency standard (Symmetricom, E4410A). The resulting error signal is processed in a proportional-integral servo controller (New Focus, LB1005), amplified (Piezomechanik GmbH, SVR 200/3), and fed to the PZT fiber stretcher in the Er:fiber oscillator. When the lock is engaged, the full-width at half-maximum (FWHM) of the 26$^{th}$ harmonic of the $f_{rep}$ is 2 Hz, measured using a spectrum analyzer with 1 Hz resolution bandwidth and a sweep time of 1.8 s. From this, we estimate that the contribution of the $f_{rep}$ lock to the linewidth of the MIR comb modes is at most 25 kHz.

Behind the OP-GaP crystal, an optical long-pass filter blocks the near-IR radiation emitted by the dual-wavelength source while transmitting the MIR idler produced in the DFG process. The latter is collimated and coupled into a multi-pass absorption cell with an absorption path length of 10 m (Thorlabs, HC10L/M-M02). The cell is connected to a vacuum pump, a pressure transducer (Leybold, CERAVAC CTR 101 N), and the sample gas supply (2.98 % $N_2O$ diluted in $N_2$, supplier: Air Liquide AB), as shown in the right corner of Fig. 1.

After the multi-pass cell, the MIR beam is re-collimated using two lenses and guided to a fast-scanning FTS. The FTS is the same as in Ref. [35], except that the beamsplitter and the detectors have been replaced to extend the operating range to 8 μm. The two out-of-phase interferometer outputs are detected by a pair of thermoelectrically-cooled HgCdTe detectors (VIGO Systems, PVI-4TE-10.6-1x1) in a balanced configuration. The comb beam radius inside the FTS varies between 2 mm at the waist (positioned ~0.5 m after the beamsplitter) and 4 mm at the detectors. Due to the relatively large divergence of the 8 μm beam, the off-axis components give rise to interference fringes that are phase shifted with respect to the interference fringes at the beam center, thus reducing the interferogram contrast. To mitigate this effect, we placed an optical aperture in front of each detector, which yielded an interferogram contrast of ~50% (compared to ~30% without the aperture). The aperture radius (~2 mm) was found empirically as a trade-off between the interferogram contrast and the power level on the detectors (~14 μW). The optical path difference in the FTS is calibrated using a 1556 nm narrow-linewidth CW laser (RIO, PLANEX), whose beam propagates on a path parallel to the MIR comb beam. The CW laser interferogram is registered using an InGaAs detector. The comb and CW laser interferograms are recorded synchronously with an analog-digital converter (National Instruments, PCI-5922) and a LabVIEW program running on a personal computer.

To acquire a spectrum, we first evacuated the cell, purged it with the sample gas, filled it to the desired total pressure, and closed the valve at the input to the cell to isolate it from the rest of the gas system. We stabilized the comb $f_{rep}$ and acquired a set of 50 interferograms. To reduce the sampling point spacing in the spectrum, we stepped $f_{rep}$ in increments of 29 Hz via tuning of the RF generator, and acquired 50 interferograms at each step. In total, we made 14 steps of $f_{rep}$ to scan each comb mode over one mode spacing ($f_{rep}$) in the optical domain. We made two measurements, at total pressures of 0.02 mbar and 0.32 mbar, with signal to noise ratio optimized for the $v_1$ fundamental band and the $v_1 + v_2 - v_2$ hot band, respectively. At 0.02 mbar, we made six $f_{rep}$ scans in alternating directions, resulting in a total of 300 interferograms for each $f_{rep}$ setting. One interferogram was acquired in 3.6 s, which yielded a total acquisition time of 4.2 h. At 0.32 mbar, we made two $f_{rep}$ scans with 50 interferograms per $f_{rep}$ step, as well as a final scan with 25 interferograms per step, yielding 125 interferograms per $f_{rep}$ setting and an acquisition time of 1.8 h. To obtain a reference spectrum for normalization, we recorded interferograms with the absorption cell evacuated and the comb locked to the first $f_{rep}$ step. For the measurement at 0.02 mbar, we recorded 150 reference interferograms before and after measuring the $N_2O$ absorption spectrum, while for the measurement at 0.32 mbar, we recorded 125 reference interferograms before the $N_2O$ measurement.

We processed the acquired data using a MATLAB code. We first resampled the comb interferograms at the zero-crossings and extrema of the CW laser interferogram and averaged the absolute values of the Fourier transforms of each set of interferograms belonging to one $f_{rep}$ step. We used the method described in Refs [36, 37] in order to match the frequency domain sampling points to the comb mode positions at each $f_{rep}$ step and thus minimize the influence of the instrumental line shape. We normalized the averaged spectrum at each $f_{rep}$ step to the averaged reference spectrum, which had been smoothened and interpolated linearly to the sampling points of the pertinent $f_{rep}$ step. Smoothening and linear interpolation was possible since the background features were broad enough to be fully resolved at the comb mode spacing. Using the Lambert-Beer law, we converted each transmission spectrum to an absorption spectrum. To remove the remaining baseline, we fit a model consisting of a sum of a simulated absorption spectrum based on the parameters from the HITRAN 2016 database [13] and a baseline, represented by the sum of a $5^{th}$ order polynomial and slowly varying sine terms (periods >6 GHz) to the absorption spectra, and then subtracted the baseline. Finally, we interleaved the averaged and baseline-corrected absorption spectra recorded at different $f_{rep}$ steps to obtain the final spectrum with a sample point spacing of 9 MHz in the optical domain.

## 3. Results and discussion

### 3.1. High-resolution spectra

Figure 2(a) shows the absorption spectrum of $N_2O$ measured at 0.02 mbar, dominated by the P and R branches of the $\nu_1$ fundamental band centered at 1285 cm$^{-1}$. A second progression of lines, one order of magnitude weaker, belongs to the $\nu_1 + \nu_2 - \nu_2$ hot band that is split into two components designated by their rotationless parity (e/f). To improve the signal-to-noise ratio (SNR) of these lines, we made a measurement at a higher pressure of 0.32 mbar. Figure 2(b) shows a vertical zoom of the hot band in this second measurement, where the lines belonging to the $\nu_1$ band are plotted in gray for clarity. In addition to the strong P and R branches of the hot band, there is a weaker Q-branch around 1290 cm$^{-1}$. Due to minor water impurities in the cell, a few water lines are visible [e.g., around 1260 cm$^{-1}$ in Fig. 2(a)], some of them accompanied by slight baseline distortions caused by water absorption in the ambient air. The maximum SNR of the lines in the $\nu_1$ band in Fig. 2(a) is ~420, while for the hot band in Fig. 2(b) it is ~200. The SNR is limited by the detector noise.

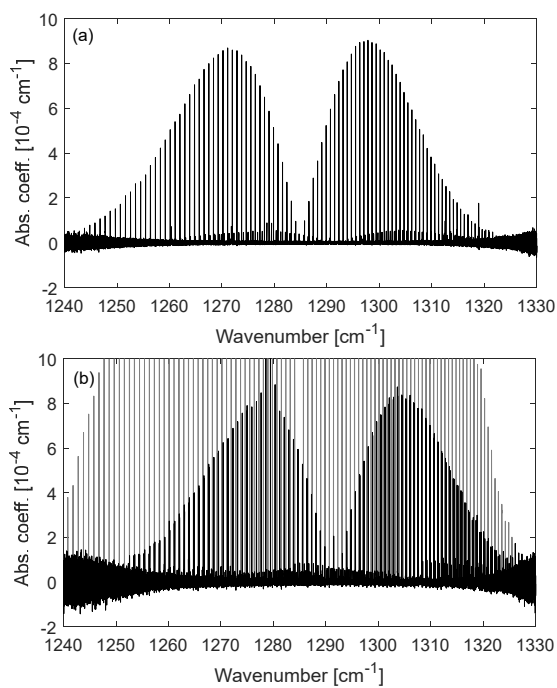

Fig. 2 Measured absorption spectra of 2.98% $N_2O$ in $N_2$ at (a) 0.02 mbar (300 averages) and (b) 0.32 mbar (125 averages). In (b), the spectrum is vertically zoomed in to show the $\nu_1 + \nu_2 - \nu_2$ hot band, while the lines belonging to the fundamental $\nu_1$ band are plotted in gray.

### 3.2. Line by line fitting

To retrieve the line center positions of the individual absorption lines, we fit Voigt line shapes to them. We applied the fit to lines with line strengths exceeding a threshold of $2 \times 10^{-20}$ cm$^{-1}$/(molecule/cm$^2$) for the $\nu_1$ band and $2 \times 10^{-21}$ cm$^{-1}$/(molecule/cm$^2$) for the hot band (based on the values from the HITRAN database [13]). These criteria translate to a minimum SNR of ~30. In the Voigt model, we fit the intensities, the Lorentzian widths, and the center frequencies, together with a linear baseline. We fixed the Doppler widths to the theoretical values calculated at 23 ºC, which are around 71 MHz (FWHM). We used a nonlinear fitting routine based on the Levenberg-Marquardt algorithm, and we chose a fitting range of ±426 MHz for each line, equal to roughly ±6 times the FWHM. Lines separated by less than 9 times the Doppler FWHM were fitted simultaneously in one window. Such cases occurred only for the hot band since the line pairs with different parity overlap in parts of the spectrum. Lines separated by less than twice the Doppler FWHM were excluded from fitting. Weak interfering lines from other $N_2O$ bands and isotopologues within the fit window were included in the fit with their parameters fixed to the HITRAN values if they exceeded 2% of the line strength threshold. Weaker interfering lines were ignored since they were below the noise floor. We kept the line fits with a quality factor (i.e., the ratio of the peak of the line and the standard deviation of the residuum) larger than 30, thereby eliminating some lines that were affected by baseline distortions, e.g., due to interfering water absorption. Figure 3 shows fits to the R(14) line of the $\nu_1$ band [Fig. 3(a)], and to a pair of lines belonging to the R(4) transition of the hot band with e- and f-parity [Fig. 3(b)]. The residuals displayed in the lower panels show no significant structure.

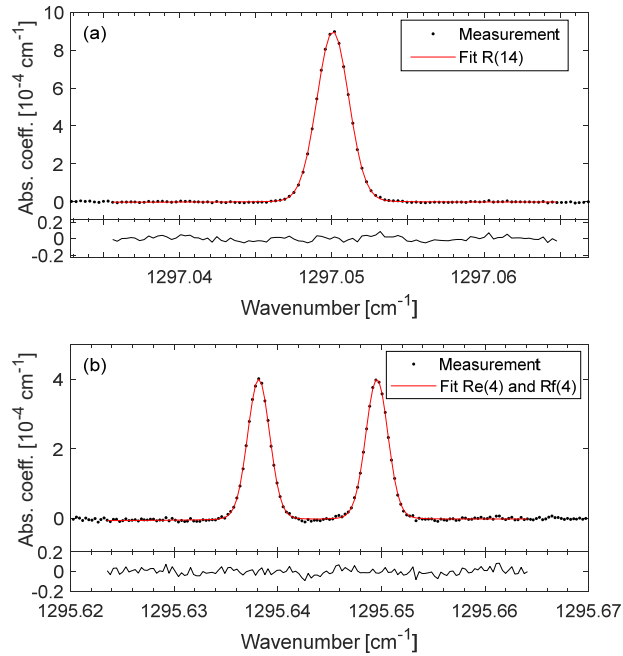

Fig. 3 Spectra of the (a) R(14) line in the $\nu_1$ fundamental band (black markers), and (b) the R(4) line of the hot band with the e-parity (left) and f-parity (right) together with fits of the Voigt profile (red curves). The lower panels show the residuals within the fitting windows.

The linewidth at the pressures used is dominated by the Doppler broadening, and the pressure broadening is expected to be ~90 kHz for the $\nu_1$ band (at 0.02 mbar) and ~1.5 MHz for the hot band (at 0.32 mbar), as calculated using the pressure broadening parameters from the HITRAN database [13]. The fitted Lorentzian FWHMs are larger by on average 4.2 MHz than the expected values, with no clear J-dependence. We assume that this discrepancy is due to an instrumental broadening, and we made a number of tests to investigate its origin. The characterization of the $f_{rep}$ locking stability described in the experimental section excludes it as a possible cause. To rule out that the broadening stemmed from the drift of the CW reference laser wavelength during the few-hour long measurement, we replaced the RIO diode laser with a CW Er:fiber laser (Koheras Adjustik E15) stabilized to an Er:fiber frequency comb, but this yielded no reduction of the broadening. We also found no dependence of the broadening on the size of the apertures in front of the detectors. We therefore suspect that the broadening is caused by the linewidth of the modes of the 8 μm idler comb, but we have no means of measuring it at this time. Since the line broadening appears symmetric, as indicated by the flat residuals in Fig. 3, we assume it does not skew the retrieved line center positions.

We identify three sources of uncertainty of the fitted line positions. The dominating component is the fit precision, which is in the range of 50-600 kHz. The second contribution stems from the uncertainty of the effective CW reference laser wavelength, which depends on the alignment and divergence of the two beams in the FTS, as well as variations of the refractive index of air that fills the FTS. Using the method described in Refs [35, 37] for determination of the reference wavelength and its effect on the retrieved line positions, we estimate its contribution to the uncertainty to be 60 kHz. The third source of uncertainty are the pressure shifts. The pressure shifts of the lines in the hot band at 0.32 mbar are on average 10 kHz (according to HITRAN) and we subtracted them from the fitted line center frequencies. We conservatively assumed the uncertainty of the pressure shift to be as large as the shift itself. The pressure shifts of the lines in the fundamental band at 0.02 mbar are lower than 3 kHz, i.e., more than one order of magnitude below the next largest contribution to the uncertainty budget (originating from the effective reference laser wavelength). We thus neglected pressure shifts in the analysis of the fundamental band.

### 3.3. Line positions of the $\nu_1$ band

The line centers retrieved by fitting to 72 lines in the $\nu_1$ fundamental band are presented in Table 1. The uncertainties (160 kHz on average) are calculated by summing in quadrature the fit precision and the contribution from the reference laser wavelength, described in the previous section. The black dots in Fig. 4(a) show a comparison of the fitted line centers to those from the HITRAN database. A clear systematic discrepancy is visible, though remaining within the HITRAN frequency uncertainty of 3 MHz. Furthermore, this systematic tendency closely follows that reported in a recent work by AlSaif *et al.* [21], using a comb-referenced external-cavity QCL, shown by the red circles. The root-mean-square discrepancy between the line centers retrieved in that study and in this work is 240 kHz.

Similarly to what was done by AlSaif *et al.*, we fit a model based on the effective ro-vibrational Hamiltonian (see Eq. (1) in [21]) to the retrieved line positions using a Levenberg-Marquardt algorithm implemented in MATLAB. We weighted the lines by the inverse squares of their uncertainties. We fit the vibrational term value $G_v$, the rotational constant $B_v$, and the quartic and sextic centrifugal distortion constants $D_v$, $H_v$ of the vibrationally excited state (fixing $L_v$ to 0) and fixed the ground-state rotational and centrifugal distortion constants to those determined by Ting *et al.* [38]. The fitted upper-state constants of the $\nu_1$ band are presented in Table 2, and compared to those obtained by AlSaif *et al.* [21]. For further comparison, we include the constants reported by Tachikawa *et al.* [39] from sub-Doppler measurements of the $\nu_3 - \nu_1$ and $\nu_3 - \nu_2$ bands performed by heterodyning two fluorescence-stabilized lasers. Except for $G_v$, the retrieved upper-state constants from our work and Ref. [21] agree within their combined 1σ uncertainties; the $G_v$ constants agree within the combined 3σ uncertainty. The agreement with Tachikawa *et al.* is noticeably worse for all constants except $B_v$. Note however that their values resulted from fits to two other bands and might be influenced by factors not present in the two other studies, in which the $\nu_1$ band was measured directly. Figure 4(b) shows a comparison of the line centers measured in this work to those calculated from the fitted spectroscopic constants (black dots). The weighted standard deviation (observed – calculated) of the 72 fitted lines is $3.2 \times 10^{-6}$ cm$^{-1}$ (96 kHz). The red circles and blue triangles show a comparison to the line positions calculated using the constants from Refs [21] and [39] respectively. Considerable disagreement with the latter is apparent mainly as a 690 kHz offset. On the other hand, the discrepancy between our model and the one from AlSaif *et al.* remains within the uncertainties of the measured line positions in the studied wavenumber range. We thus provide an independent support of the accuracy of the line positions and constants reported by AlSaif *et al.* [21], using a different measurement technique.

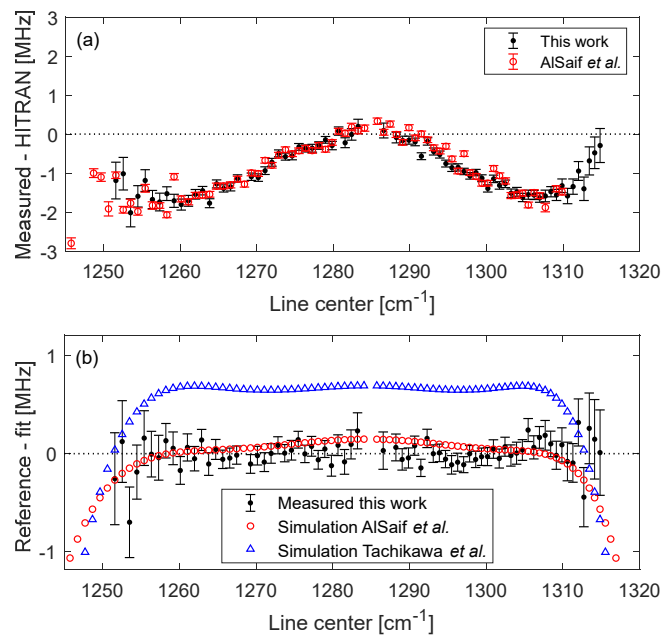

Fig. 4 (a) The line positions of the $\nu_1$ fundamental band from this work (black dots) and from Ref. [21] (red circles) relative to those from the HITRAN database [13]. (b) The line positions obtained by line-by-line fitting (black dots) relative to the simulation based on the upper state constants from this work. The red circles and blue triangles indicate line positions simulated using constants from Ref. [21] and Ref. [39] relative to the simulation from our work.

Table 1. Line centers of the $\nu_1$ band retrieved by line-by-line fitting.

| Transition | Line center [cm$^{-1}$] | Transition | Line center [cm$^{-1}$] |
| --- | --- | --- | --- |
| P(37) | 1251.59987(2) | R(3) | 1288.220267(5) |
| P(36) | 1252.56081(2) | R(4) | 1289.040685(4) |
| P(35) | 1253.51840(2) | R(5) | 1289.857575(4) |
| P(34) | 1254.47274(1) | R(6) | 1290.670933(4) |
| P(33) | 1255.42375(1) | R(7) | 1291.480741(3) |
| P(32) | 1256.371424(8) | R(8) | 1292.287024(4) |
| P(31) | 1257.315772(8) | R(9) | 1293.089745(4) |
| P(30) | 1258.256789(6) | R(10) | 1293.888923(3) |
| P(29) | 1259.194453(6) | R(11) | 1294.684545(3) |
| P(28) | 1260.128760(5) | R(12) | 1295.476611(4) |
| P(27) | 1261.059723(5) | R(13) | 1296.265122(4) |
| P(26) | 1261.987308(5) | R(14) | 1297.050068(3) |
| P(25) | 1262.911532(4) | R(15) | 1297.831454(3) |
| P(24) | 1263.832361(4) | R(16) | 1298.609267(3) |
| P(23) | 1264.749817(4) | R(17) | 1299.383512(3) |
| P(22) | 1265.663873(4) | R(18) | 1300.154183(4) |
| P(21) | 1266.574535(4) | R(19) | 1300.921282(4) |
| P(20) | 1267.481792(4) | R(20) | 1301.684796(4) |
| P(18) | 1269.286064(4) | R(21) | 1302.444737(4) |
| P(17) | 1270.183073(4) | R(22) | 1303.201089(4) |
| P(16) | 1271.076649(4) | R(23) | 1303.953861(4) |
| P(15) | 1271.966796(4) | R(24) | 1304.703046(4) |
| P(14) | 1272.853504(3) | R(25) | 1305.448648(4) |
| P(13) | 1273.736761(3) | R(26) | 1306.190648(4) |
| P(12) | 1274.616571(4) | R(27) | 1306.929065(4) |
| P(11) | 1275.492929(3) | R(28) | 1307.663887(6) |
| P(10) | 1276.365818(3) | R(29) | 1308.395111(5) |
| P(9) | 1277.235248(4) | R(30) | 1309.122738(5) |
| P(8) | 1278.101200(4) | R(31) | 1309.846776(6) |
| P(7) | 1278.963685(3) | R(32) | 1310.567207(7) |
| P(6) | 1279.822680(4) | R(33) | 1311.284045(7) |
| P(5) | 1280.678203(4) | R(34) | 1311.997298(9) |
| P(4) | 1281.530223(5) | R(35) | 1312.70691(2) |
| P(3) | 1282.378760(5) | R(36) | 1313.41298(2) |
| P(2) | 1283.223797(7) | R(37) | 1314.11541(2) |
| R(1) | 1286.568843(7) | R(38) | 1314.81425(2) |

Table 2. The vibrational term value $G_v$, rotational constant $B_v$, and centrifugal distortion constants $D_v$, $H_v$ calculated in this work and in Refs [21]* and [39]. Units are cm$^{-1}$.

| $G_v$ | $B_v$ | $D_v \times 10^6$ | $H_v \times 10^{12}$ | Source |
|---|---|---|---|---|
| 1284.9033391(12) | 0.417255089(10) | 0.172623(19) | 0.147(11) | This work |
| 1284.9033441(13) | 0.417255073(10) | 0.172597(18) | 0.1307(86) | AlSaif *et al.* [21] |
| 1284.9033623(36) | 0.41725507388(69) | 0.17257676(67) | 0.11164(23) | Tachikawa *et al.* [39] |

### 3.4. Line positions of the $v_1 + v_2 - v_2$ hot band

The positions of the 112 fitted lines of the P and R branches of the hot band are given in Table 3. The uncertainties include all three contributions described in Section 3.2 summed in quadrature. Their mean is 230 kHz. Figure 5(a) shows the center frequencies compared to those listed in the HITRAN database (black dots) and reported by Maki and Wells [16] (red cicles), where the plotted uncertainty is from our work only. The missing lines between 1277 cm$^{-1}$ and 1283 cm$^{-1}$ were excluded from the fitting procedure due to the overlap of the e- and f-parity lines. A second gap between 1288 cm$^{-1}$ and 1294 cm$^{-1}$ is due to too low SNR close to the band center. The discrepancies with respect to HITRAN have a slight *J*-dependence; the agreement is excellent at the band center, and worse at higher *J* numbers, but all deviations are within the HITRAN uncertainty of 3 MHz. The discrepancy with respect to Maki and Wells is larger and has a clearer *J*-dependence, but it is within the 3σ uncertainty (2.3 MHz) reported in their work.

We modeled the hot band with the same effective Hamiltonian as the fundamental band, using a common vibrational term value $G_v$ for the e- and f-component, and independent $B_v$, $D_v$ and $H_v$ constants. We restricted the fit to the $G_v$, $B_v$ and $D_v$ constants of the upper states, while fixing the upper state $H_v$ constants as well as all the lower state constants. (Floating the $H_v$ constants resulted in uncertainties comparable to their absolute values.) We took the values for the fixed constants as well as the initial values for the fitted ones from Ref. [12], which is the source of the line positions listed in HITRAN. We applied the fit to all 112 lines weighted by the inverse squares of their uncertainties. Figure 5(b) shows the measured line positions relative to the fit (red solid circles and blue solid squares), where the weighted standard deviation (observed – calculated) is $7.0 \times 10^{-6}$ cm$^{-1}$ (210 kHz). Also indicated are the line positions calculated using the constants from Toth [12] for the e- and f-parity lines (red open circles and blue open squares, respectively). Table 4 shows the band origins, calculated as $v_0 = G_v' - G_v''$, where $G_v'$ and $G_v''$ are the vibrational term values of the upper and lower states, and the $B_v$ and $D_v$ constants obtained in this work compared to those from Ref. [12]. The uncertainty in the band origin, $v_0$, from Toth was calculated as the in-quadrature sum of the uncertainties of $G_v'$ and $G_v''$ reported in Ref. [10] (both equal to $1 \times 10^{-5}$ cm$^{-1}$). We note the agreement with Toth in the band origin and the improvement of its precision by one order of magnitude. The significance of the discrepancy in the remaining constants is difficult to evaluate, since no uncertainties are reported on these constants in Ref. [12]. The accuracy of the model also depends on the fixed lower state constants, which are presumably known with lower certainty than their ground state counterparts.

---

* We note that the vibrational term value $G_v$ in Table 3 of Ref. [21] is stated as 1284.9033344(13) cm$^{-1}$. The value presented in this table was obtained through private communication with Marco Marangoni.

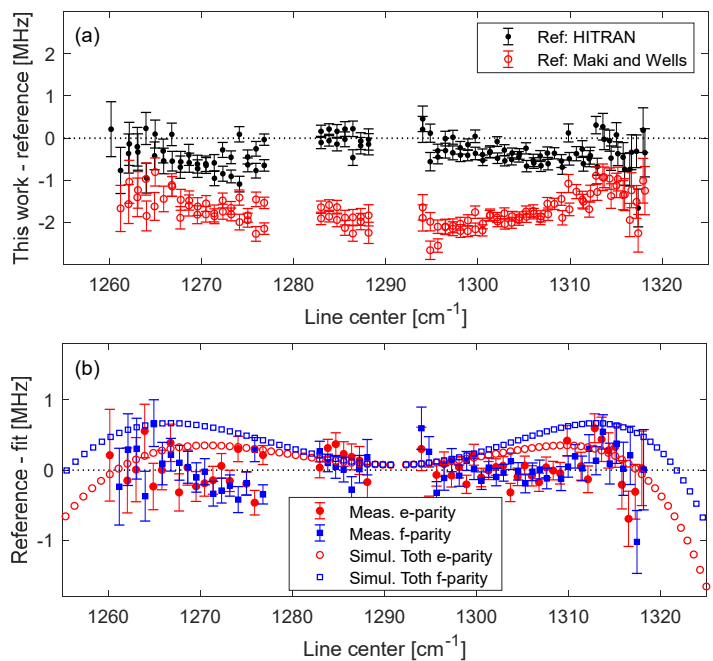

Fig. 5 (a) The line positions of the $\nu_1 + \nu_2 - \nu_2$ hot band from this work relative to the positions from the HITRAN database [13] (black dot) and from Maki and Wells [16] (red circles). The uncertainties in the line positions are from our work only. (b) The positions of the e-parity (red solid circles) and f-parity (blue solid squares) lines obtained by line-by-line fitting relative to the simulation based on the upper-state constants from our work. The red open circles and blue open squares show line positions calculated using constants from Toth [12] relative to the simulation from our work.

Table 3. Line centers of the $\nu_1 + \nu_2 - \nu_2$ bands retrieved by line-by-line fitting.

| Transition | Line center [cm$^{-1}$] | Transition | Line center [cm$^{-1}$] |
| --- | --- | --- | --- |
| Pe(35) | 1260.15238(3) | Re(5) | 1296.455900(6) |
| Pf(34) | 1261.18600(2) | Rf(5) | 1296.470290(6) |
| Pe(33) | 1262.05259(2) | Re(6) | 1297.270178(6) |
| Pf(33) | 1262.12783(2) | Rf(6) | 1297.287801(6) |
| Pe(32) | 1262.99785(2) | Re(7) | 1298.081012(6) |
| Pf(32) | 1263.06668(2) | Rf(7) | 1298.102077(6) |
| Pe(31) | 1263.93989(2) | Re(8) | 1298.888376(5) |
| Pf(31) | 1264.00255(2) | Rf(8) | 1298.913145(5) |
| Pe(30) | 1264.87863(2) | Re(9) | 1299.692301(5) |
| Pf(30) | 1264.93550(2) | Rf(9) | 1299.720978(5) |
| Pe(29) | 1265.81414(1) | Re(10) | 1300.492737(5) |
| Pf(29) | 1265.86542(2) | Rf(10) | 1300.525583(5) |
| Pe(28) | 1266.746393(9) | Re(11) | 1301.289719(5) |
| Pf(28) | 1266.792372(8) | Rf(11) | 1301.326969(5) |
| Pe(27) | 1267.675337(9) | Re(12) | 1302.083227(4) |
| Pf(27) | 1267.716311(9) | Rf(12) | 1302.125114(5) |
| Pe(26) | 1268.601037(9) | Re(13) | 1302.873261(5) |
| Pf(26) | 1268.637241(8) | Rf(13) | 1302.920032(4) |
| Pe(25) | 1269.523430(8) | Re(14) | 1303.659803(5) |
| Pf(25) | 1269.555146(7) | Rf(14) | 1303.711718(4) |
| Pe(24) | 1270.442540(7) | Re(15) | 1304.442885(5) |
| Pf(24) | 1270.470028(7) | Rf(15) | 1304.500158(5) |
| Pe(23) | 1271.358350(7) | Re(16) | 1305.222482(5) |
| Pf(23) | 1271.381858(6) | Rf(16) | 1305.285358(5) |
| Pe(22) | 1272.270861(6) | Re(17) | 1305.998582(5) |
| Pf(22) | 1272.290652(6) | Rf(17) | 1306.067330(5) |
| Pe(21) | 1273.180045(6) | Re(18) | 1306.771196(5) |
| Pf(21) | 1273.196390(6) | Rf(18) | 1306.846058(5) |
| Pe(20) | 1274.085933(7) | Re(19) | 1307.540328(5) |
| Pf(20) | 1274.099054(7) | Rf(19) | 1307.621540(5) |
| Pe(19) | 1274.988465(6) | Re(20) | 1308.305958(5) |
| Pf(19) | 1274.998659(6) | Re(21) | 1309.068092(6) |
| Pe(18) | 1275.887674(6) | Rf(21) | 1309.162787(6) |
| Pf(18) | 1275.895192(6) | Re(22) | 1309.826744(8) |
| Pe(17) | 1276.783579(5) | Rf(22) | 1309.928554(6) |
| Pf(17) | 1276.788598(5) | Rf(23) | 1310.691078(6) |
| Pf(10) | 1282.955535(5) | Re(24) | 1311.333501(7) |
| Pe(10) | 1282.960857(5) | Rf(24) | 1311.450350(6) |
| Pf(9) | 1283.823998(5) | Re(25) | 1312.081624(7) |
| Pe(9) | 1283.829857(5) | Rf(25) | 1312.206396(7) |
| Pf(8) | 1284.689315(6) | Re(26) | 1312.826270(7) |
| Pe(8) | 1284.695465(6) | Rf(26) | 1312.959187(7) |
| Pf(7) | 1285.551479(6) | Re(27) | 1313.56738(2) |
| Pe(7) | 1285.557677(6) | Rf(27) | 1313.708759(9) |
| Pf(6) | 1286.410475(7) | Re(28) | 1314.304978(9) |
| Pe(6) | 1286.416497(7) | Rf(28) | 1314.45506(1) |
| Pf(5) | 1287.266324(7) | Re(29) | 1315.03907(1) |
| Pe(5) | 1287.271918(7) | Rf(29) | 1315.19816(1) |
| Pf(4) | 1288.118999(9) | Re(30) | 1315.76964(2) |
| Pe(4) | 1288.123925(9) | Rf(30) | 1315.93800(2) |
| Re(2) | 1293.99237(2) | Re(31) | 1316.49669(2) |
| Rf(2) | 1293.99853(2) | Rf(31) | 1316.67462(2) |
| Re(3) | 1294.816961(8) | Re(32) | 1317.22025(2) |
| Rf(3) | 1294.825654(8) | Rf(32) | 1317.40796(2) |
| Re(4) | 1295.638160(6) | Re(33) | 1317.94030(2) |
| Rf(4) | 1295.649565(6) | Rf(33) | 1318.13816(2) |

Table 4. The band origin $v_0$ and the upper state constants $B_v$ and $D_v$ of the hot band e- and f-components calculated in this work compared to those from Ref. [12]. Units are cm$^{-1}$.

| $v_0$ | $B_v$ (e) | $B_v$ (f) | $D_v$ (e) × 10$^6$ | $D_v$ (f) × 10$^6$ | Source |
|---|---|---|---|---|---|
| 1291.497868(2) | 0.417464643(8) | 0.418372943(8) | 0.17482(1) | 0.17192(1) | This work |
| 1291.49787(2) | 0.417464677 | 0.418372995 | 0.1748503 | 0.1719561 | Toth [12] |

## 4. Conclusions

We demonstrated a Fourier transform spectrometer based on a compact fiber-based DFG optical frequency comb source and a multi-pass cell capable of measuring low-pressure spectra with center frequency precision of the order of 100 kHz in the atmospheric window around 8 μm. The DFG comb source is offset-frequency-free, which implies that an RF lock of the repetition rate is sufficient to achieve absolute stability of the comb mode frequencies. This, in combination with the FTS with sub-nominal resolution sampling-interleaving scheme, provides spectra with a calibrated frequency axis. We verified the accuracy of the retrieved line positions of the $v_1$ fundamental band of $N_2O$ by demonstrating excellent agreement with the results of an independent high-precision measurement by AlSaif *et al*. [21] using an external-cavity QCL referenced to a frequency comb. We reached a comparable precision in the retrieved line positions using a system with reduced experimental complexity. We also retrieved line positions of the $v_1 + v_2 - v_2$ hot band with up to one order of magnitude improved precision compared to previous studies [5, 12, 16]. This allowed us to determine the band origin with an uncertainty reduced by one order of magnitude compared to the FTIR-based work of Toth [12]. Longer averaging times will further increase the SNR and hence the precision of the retrieved line positions. Furthermore, the wavelength coverage of the DFG comb can be expanded within the 7-9 μm range by changing the poling period of the crystal and tuning the soliton signal [25]. Our work opens up precision measurements of entire bands of various molecules of interest in atmospheric science and astrophysics, such as methane, ammonia, sulphur dioxide or methanol, in this fingerprint spectral region.

**Acknowledgments**

The authors thank Isak Silander and Vinicius Silva de Oliveira for assistance in setting up the comb-referenced CW Er:fiber laser, Thorlabs Sweden for the loan of the multi-pass cell, and Marco Marangoni for helpful discussions concerning the results from Ref. [21]. This project is financed by the Knut and Alice Wallenberg Foundation (KAW 2015.0159), the Swedish Research Council (2016-03593), and the Foundation for Polish Science (First TEAM/2017-4/39).

**References**

[1] Montzka SA, Dlugokencky EJ, Butler JH. Non-$CO_2$ greenhouse gases and climate change. Nature. 2011;476:43-50.

[2] Ravishankara AR, Daniel JS, Portmann RW. Nitrous oxide ($N_2O$): The dominant ozone-depleting substance emitted in the 21st century. Science. 2009;326:123-5.

[3] Meure CM, Etheridge D, Trudinger C, Steele P, Langenfelds R, van Ommen T, et al. Law dome $CO_2$, $CH_4$ and $N_2O$ ice core records extended to 2000 years BP. Geophys Res Lett. 2006;33:L14810.

[4] Griffis TJ, Chen Z, Baker JM, Wood JD, Millet DB, Lee X, et al. Nitrous oxide emissions are enhanced in a warmer and wetter world. Proc Natl Acad Sci USA. 2017;114:12081-5.

[5] Guelachvili G. Absolute $N_2O$ wavenumbers between 1118 and 1343 cm$^{-1}$ by Fourier-transform spectroscopy. Can J Phys. 1982;60:1334-47.

[6] Levy A, Lacome N, Guelachvili G. Measurement of $N_2O$ line strengths from high-resolution Fourier-transform spectra. J Mol Spectrosc. 1984;103:160-75.

[7] Lacome N, Levy A, Guelachvili G. Fourier-transform measurement of self-broadening, $N_2$-broadening, and $O_2$-broadening of $N_2O$ lines – temperature-dependence of linewidths. Appl Opt. 1984;23:425-35.

[8] Toth RA. Line strengths of $N_2O$ in the 1120–1440-cm$^{-1}$ region. Appl Opt. 1984;23:1825-34.

[9] Toth RA. Frequencies of $N_2O$ in the 1100-cm$^{-1}$ to 1440-cm$^{-1}$ region. J Opt Soc Am B. 1986;3:1263-81.

[10] Toth RA. Line-frequency measurements and analysis of $N_2O$ between 900 and 4700 $cm^{-1}$. Appl Opt. 1991;30:5289-315.

[11] Toth RA. Line strengths (900–3600 $cm^{-1}$), self-broadened linewidths, and frequency-shifts (1800–2360 $cm^{-1}$) of $N_2O$. Appl Opt. 1993;32:7326-65.

[12] Toth RA. Linelist of $N_2O$ parameters from 500 to 7500 $cm^{-1}$. http://mark4sun.jpl.nasa.gov/n2o.html.

[13] Gordon IE, Rothman LS, Hill C, Kochanov RV, Tan Y, Bernath PF, et al. The HITRAN2016 molecular spectroscopic database. J Quant Spectrosc Radiat Transf. 2017;203:3-69.

[14] Wang CY, Liu AW, Perevalov VI, Tashkun SA, Song KF, Hu SM. High-resolution infrared spectroscopy of $^{14}N^{15}N^{16}O$ and $^{15}N^{14}N^{16}O$ in the 1200–3500 $cm^{-1}$ region. J Mol Spectrosc. 2009;257:94-104.

[15] Varanasi P, Chudamani S. Line strength measurements in the $\nu_1$-fundamental band of $^{14}N_2^{16}O$ using a tunable diode-laser. J Quant Spectrosc Radiat Transf. 1989;41:359-62.

[16] Maki AG, Wells JS. New wavenumber calibration tables from heterodyne frequency measurements. J Res Natl Inst Stan. 1992;97:409-70.

[17] Grossel A, Zeninari V, Parvitte B, Joly L, Courtois D, Durry G. Quantum cascade laser spectroscopy of $N_2O$ in the 7.9 μm region for the in situ monitoring of the atmosphere. J Quant Spectrosc Radiat Transf. 2008;109:1845-55.

[18] Tonokura K, Takahashi R. Pressure broadening of the $\nu_1$ band of nitrous oxide by carbon dioxide. Chem Lett. 2016;45:95-7.

[19] Tasinato N, Hay KG, Langford N, Duxbury G, Wilson D. Time dependent measurements of nitrous oxide and carbon dioxide collisional relaxation processes by a frequency down-chirped quantum cascade laser: Rapid passage signals and the time dependence of collisional processes. J Chem Phys. 2010;132:164301.

[20] Lamperti M, AlSaif B, Gatti D, Fermann M, Laporta P, Farooq A, et al. Absolute spectroscopy near 7.8 μm with a comb-locked extended-cavity quantum-cascade-laser. Scientific Reports. 2018;8:1292.

[21] AlSaif B, Lamperti M, Gatti D, Laporta P, Fermann M, Farooq A, et al. High accuracy line positions of the $\nu_1$ fundamental band of $^{14}N_2^{16}O$. J Quant Spectrosc Radiat Transf. 2018;211:172-8.

[22] AlSaif B, Gatti D, Lamperti M, Laporta P, Farooq A, Marangoni M. Comb-calibrated sub-Doppler spectroscopy with an external-cavity quantum cascade laser at 7.7 μm. Opt Express. 2019;27:23785-90.

[23] Gambetta A, Coluccelli N, Cassinerio M, Gatti D, Laporta P, Galzerano G, et al. Milliwatt-level frequency combs in the 8–14 μm range via difference frequency generation from an Er:fiber oscillator. Opt Lett. 2013;38:1155-7.

[24] Lee KF, Hensley CJ, Schunemann PG, Fermann ME. Midinfrared frequency comb by difference frequency of erbium and thulium fiber lasers in orientation-patterned gallium phosphide. Opt Express. 2017;25:17411-6.

[25] Krzempek K, Tomaszewska D, Gluszek A, Martynkien T, Mergo P, Sotor J, et al. Stabilized all-fiber source for generation of tunable broadband $f_{CEO}$-free mid-IR frequency comb in the 7–9 μm range. Opt Express. 2019;27:37435-45.

[26] Timmers H, Kowligy A, Lind A, Cruz FC, Nader N, Silfies M, et al. Molecular fingerprinting with bright, broadband infrared frequency combs. Optica. 2018;5:727-32.

[27] Vasilyev S, Moskalev I, Smolski V, Peppers J, Mirov M, Muraviev A, et al. Multi-octave visible to long-wave IR femtosecond continuum generated in Cr:ZnS-GaSe tandem. Opt Express. 2019;27:16405-13.

[28] Changala PB, Weichman ML, Lee KF, Fermann ME, Ye J. Rovibrational quantum state resolution of the $C_{60}$ fullerene. Science. 2019;363:49-54.

[29] Maidment L, Schunemann PG, Reid DT. Molecular fingerprint-region spectroscopy from 5 to 12 μm using an orientation-patterned gallium phosphide optical parametric oscillator. Opt Lett. 2016;41:4261-4.

[30] Kara O, Maidment L, Gardiner T, Schunemann PG, Reid DT. Dual-comb spectroscopy in the spectral fingerprint region using OPGaP optical parametric oscillators. Opt Express. 2017;25:32713-21.

[31] Iwakuni K, Bui TQ, Niedermeyer JF, Sukegawa T, Ye J. Comb-resolved spectroscopy with immersion grating in long-wave infrared. Opt Express. 2019;27:1911-21.

[32] Hugi A, Villares G, Blaser S, Liu HC, Faist J. Mid-infrared frequency comb based on a quantum cascade laser. Nature. 2012;492:229-33.

[33] Gianella M, Nataraj A, Tuzson B, Jouy P, Kapsalidis F, Beck M, et al. High-resolution and gapless dual comb spectroscopy with current-tuned quantum cascade lasers. Opt Express. 2020;28:6197-208.

[34] Consolino L, Nafa M, Cappelli F, Garrasi K, Mezzapesa FP, Li LH, et al. Fully phase-stabilized quantum cascade laser frequency comb. Nat Commun. 2019;10:2938.

[35] Sadiek I, Hjältén A, Vieira FS, Lu C, Stuhr M, Foltynowicz A. Line positions and intensities of the $\nu_4$ band of methyl iodide using mid-infrared optical frequency comb Fourier transform spectroscopy. J Quant Spectrosc Radiat Transf. 2020;255:107263.

[36] Maslowski P, Lee KF, Johansson AC, Khodabakhsh A, Kowzan G, Rutkowski L, et al. Surpassing the path-limited resolution of Fourier-transform spectrometry with frequency combs. Phys Rev A. 2016;93:021802.

[37] Rutkowski L, Maslowski P, Johansson AC, Khodabakhsh A, Foltynowicz A. Optical frequency comb Fourier transform spectroscopy with sub-nominal resolution and precision beyond the Voigt profile. J Quant Spectrosc Radiat Transf. 2018;204:63-73.

[38] Ting WJ, Chang CH, Chen SE, Chen HC, Shy JT, Drouin BJ, et al. Precision frequency measurement of $N_2O$ transitions near 4.5 μm and above 150 μm. J Opt Soc Am B. 2014;31:1954-63.

[39] Tachikawa M, Evenson KM, Zink LR, Maki AG. Frequency measurements of 9- and 10-μm $N_2O$ laser transitions. IEEE J Quantum Elect. 1996;32:1732-6.